\documentclass[conference]{IEEEtran}
\IEEEoverridecommandlockouts
\usepackage{cite}
\usepackage{amsmath,amssymb,amsfonts}
\usepackage{algorithmic}
\usepackage{graphicx}
\usepackage{textcomp}
\usepackage{xcolor}
\usepackage{comment}
\usepackage{multirow}
\usepackage{booktabs}
\usepackage{array}
\usepackage{booktabs}
\usepackage{caption}
\usepackage[skins]{tcolorbox}
\usepackage{listings}
\usepackage{empheq, esvect}

\def\BibTeX{{\rm B\kern-.05em{\sc i\kern-.025em b}\kern-.08em
    T\kern-.1667em\lower.7ex\hbox{E}\kern-.125emX}}
\begin{document}

\title{PENTACET data -  23 Million Contextual Code Comments and 250,000 SATD comments 
\thanks{We thank Academy of Finland (grant ID 328058) for the financial support.}
}

\author{\IEEEauthorblockN{Murali Sridharan}
\IEEEauthorblockA{\textit{M3S, University of Oulu} \\
Oulu, Finland \\
murali.sridharan@oulu.fi}
\and
\IEEEauthorblockN{Leevi Rantala}
\IEEEauthorblockA{
\textit{M3S, University of Oulu}\\
Oulu, Finland \\
leevi.rantala@oulu.fi}
\and
\IEEEauthorblockN{Mika M{\"a}ntyl{\"a}}
\IEEEauthorblockA{
\textit{M3S, University of Oulu}\\
Oulu, Finland \\
mika.mantyla@oulu.fi}

}

\maketitle

\begin{abstract}
Most Self-Admitted Technical Debt (SATD) research utilizes explicit SATD features such as `TODO' and `FIXME' for SATD detection. A closer look reveals several SATD research uses simple SATD (`Easy to Find') code comments without contextual data (preceding and succeeding source code context). This work addresses this gap through PENTACET (or 5C dataset) data. PENTACET is a large Curated Contextual Code Comments per Contributor and the most extensive SATD data. We mine 9,096 Open Source Software Java projects totaling over 400 million LOC. 
The outcome is a dataset with 23 million code comments, preceding and succeeding source code context for each comment, and more than 250,000 SATD comments, including both `Easy to Find' and `Hard to Find' SATD.  We believe PENTACET data will further SATD research using Artificial Intelligence techniques.
 \end{abstract}


\section{Introduction}

Technical Debt (TD) refers to the cost of rework for having resorted to quicker but sub-optimal solutions in software development~\cite{cunningham1992wycash}.
Since 1990, multiple research efforts have investigated TD detection, prioritization, and management techniques. Our focus is on the type of technical debt that is detected through the source code comments left by software developers, called Self-Admitted Technical Debt (SATD)~\cite{potdar2014exploratory}. `TODO' and `FIXME' are tags\footnote[1]{https://rules.sonarsource.com/java/type/Code Smell?search=tags}used by SonarQube\cite{sonarqube23}, one of the widely used static code analysis tools\cite{vassallo2018continuous} in Continuous Integration/Deployment environment, for technical debt detection through code comments among a host of other code smells. A recent study~\cite{yu2020identifying} characterized SATD as `Easy to Find' (can be detected using simple features such as `TODO', `FIXME', `XXX', etc.,)  and `Hard to Find' (typically refers to multi-word features that do not explicitly indicate SATD, e.g., `went horribly wrong'). Identifying and characterizing the `Hard to Find' SATD will benefit software companies with improved SATD detection as a recent survey finds similar SATD comments between Open Source Software (OSS) and industry~\cite{zampetti2021self}. Our dataset is the first attempt to characterize the 'Hard to Find' SATD as proposed in \cite{yu2020identifying}. 

Several research efforts \cite{huang2018identifying},\cite{ren2019neural},\cite{guo2021far} for SATD detection utilize Maldonado's dataset~\cite{maldonado2015detecting}. It utilizes the 64 SATD patterns identified in \cite{potdar2014exploratory}. This dataset is limited to 10 OSS projects, and the dataset ails with many duplicates, as highlighted in~\cite{sridharan2021data}. Furthermore, past findings from mining  Github~\cite{cosentino2016findings} source code repositories highlight the importance of using probabilistic sampling for dataset construction instead of non-probabilistic sampling in~\cite{maldonado2015detecting,guo2021far} to enable the improved generalization of the findings.

On the human factors front, TD, apart from impacting software productivity adversely also hampers software developers' morale\cite{besker2020influence}. It is one of the vital human factors essential in delivering high-quality software~\cite{guveyi2020human}.
While multiple studies investigate the impact of developer team size (one of the crucial factors of software productivity) on attributes both internal \cite{kochhar2013empirical,youssef2015impact} and external to the software
\cite{vishnubhotla2021understanding,tyagi2022empirically}, the role of the developer team size on SATD and code commenting practices are yet to be explored. 

Motivated by these factors and inspired by the study that attempts to understand source code comments at large scale~\cite{he2019understanding}, we present PENTACET (or 5C dataset) based on well-defined curation rules. It consists of multiple clusters of bi-directional contextual (i.e., source code preceding and succeeding a code comment) source code comments mined from 9,096 OSS repositories. 

We group the clusters of contextual source code comments based on contributor count (developer team size), offering researchers leverage to associate and investigate  
the source code comments of the repository with the team size (referring to the developer with commit access to an OSS repository). 
We utilize SoCCMiner \cite{sridharan2022soccminer} to extract the bi-directional code context of a comment. We query the official GitHub REST API\footnote[2]{https://api.github.com/repos/OWNER/REPO/contributors} to fetch the contributor/s of a repository.

\textbf{Contributions: }Our contributions in this dataset are multi-fold:
\begin{enumerate}
\item first attempt to identify and characterize `Hard to Find' SATD.
\item first big SATD dataset with more than 250,000 SATD comments.
\item first dataset to capture the bi-directional source code context of SATD comments.
\item first dataset to link team size with the contextual source code comments and SATD, to the best of our knowledge.
\end{enumerate}

\textbf{Dataset Availability: }Our dataset\footnote[3]{http://dx.doi.org/10.5281/zenodo.7757462} is made available in the open forum\cite{pentacet23} to foster a host of deep learning research involving source code comments and SATD. The applications are plentiful from modeling to behavioral and predictive analytics.

\section{Methodology}
\begin{figure}[h]
  \centering
  \includegraphics[width=0.45\textwidth]{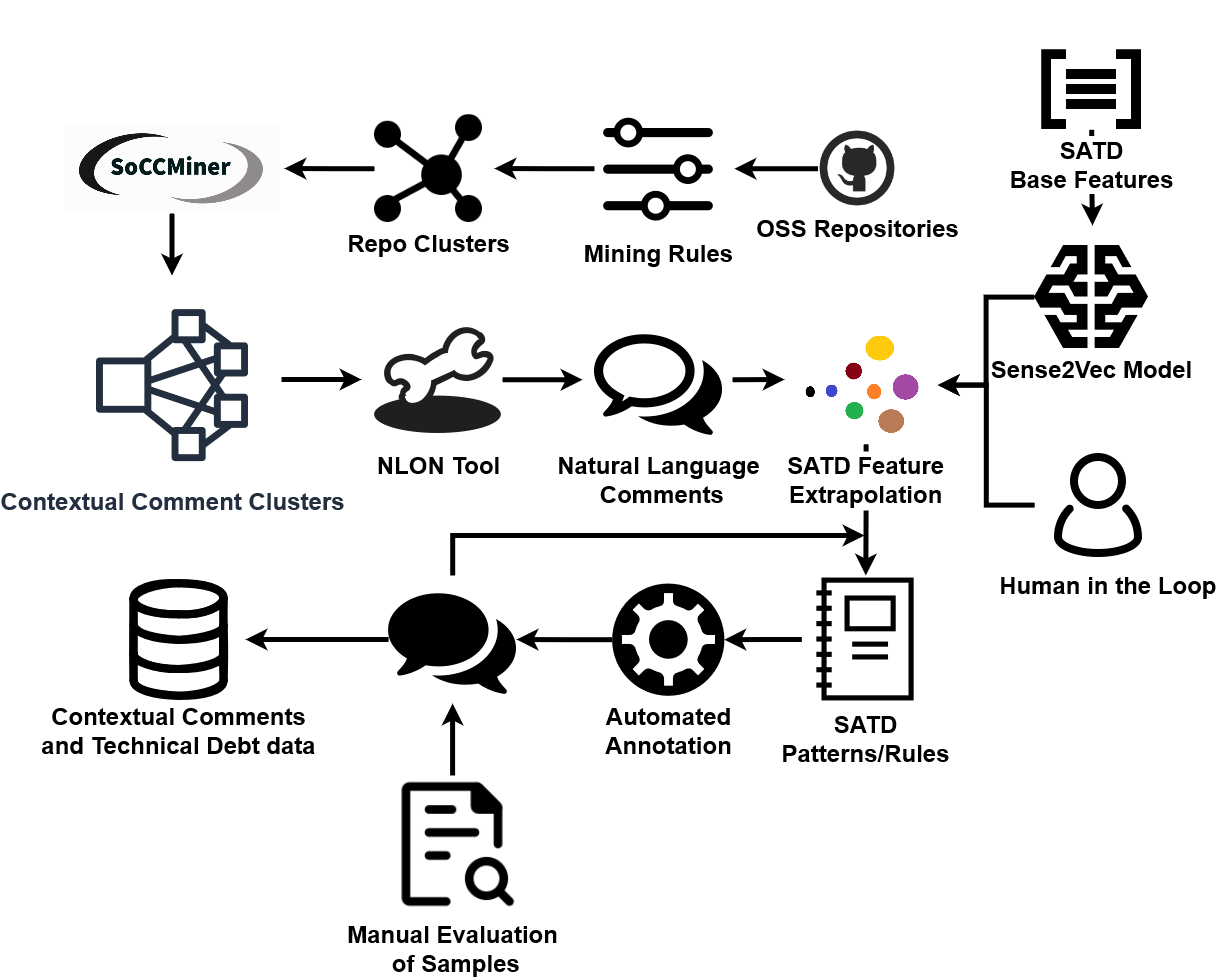}
  \caption{PENTACET - Data Construction Pipeline}
  \label{overview}
\end{figure}

The PENTACET data construction pipeline is a multi-step process. Figure~\ref{overview} depicts the methodology for creating PENTACET data. 

\textbf{Mining Rules:} The mining rules are tailored to mitigate the risks of mining OSS repositories (repos) identified in~\cite{kalliamvakou2016depth}. At first, we filter the ``Java'' OSS repositories with at least 10 stars and, have at least one code push in the past three years since January 01, 2020. It resulted in the initial repository count of 34,545. Then, to exclude non-english repositories, we apply langid\footnote[4]{https://github.com/saffsd/langid.py} tool on the repository description text. ``langid'' is a python-based language identification stand-alone tool~\cite{lui-baldwin-2011-cross}. It identified 24,718 English repositories from 34,545 repositories. Of the 24,718 repositories, we identified 3,101 repositories without any description text. Such repositories without description are excluded resulting in 21,617 repositories for the next filtering criteria. 

To eliminate the repositories that have educational materials (such as course notes, books, assignments and course materials) and inactive repositories (i.e, repositories that are not maintained and explicitly acknowledged by repository owners), the first author went through the repository (repo) description of the filtered repositories manually and identified 225 (e.g., `tutorial', `programming\_assignments', `sample repo', `demonstration project', `textbook', `curriculum', etc.,) patterns for identifying repositories containing explicit educational materials and 31 patterns (e.g., `archived', `unmaintained', `outdated', `deprecated', `no longer supported', `obsolete', `discontinued', etc.,) to identify explicit non-maintenance repositories respectively. Among the 21,617 repositories, 6,813 are identified as educational related and 244 repositories are identified as non-maintenance repositories, respectively and are excluded. It is followed by filtering repositories (a total of 202 repositories) with repository names that indicate educational or non-maintenance related repositories. For example, `AndroidDemoProjects', `Azure-Samples', `TFLiteFaceExample', etc., Now, there are 14,358 repositories for further processing. Lastly, the repositories (3,718) that are archived and without any license are removed, with 10,640 repositories progressing to next curation. Finally, to capture the active repositories, we order the repositories based on the number of repository commits~\cite{kalliamvakou2016depth} since January 01, 2020 in descending order and consider only those repositories that have atleast 3 commits since January 01, 2020, assuming atleast one commit a year. This leaves us 9,111 repositories. After removing 15 invalid repositories with `.javascript' files instead of `.java', a total of 9,096 repositories move to the next stage in the dataset construction pipeline.



\textbf{Repository Clusters:} Each OSS repository in GitHub has \textbf{`Contributors'} attribute comprising both core and co-developers \cite{matragkas2014analysing}. We cluster the curated repositories based on \textbf{`Contributor Count'} that ranges between 1 and 30 for the 9,096 repositories. We split the repositories into 4 clusters. The number of clusters are purely for organizational convenience, while any new clusters can be formed or rearranged by querying PENTACET database using `CONTRIBUTOR\_COUNT' from `PROJECT\_MAIN' table as discussed in Section \ref{dm}.  Cluster 1 (C1) contains repositories of single contributor, most likely the author of the repository. We group repositories with contributor count 2 and 3 in Cluster 2 (C2), contributor count 4 till 10 in Cluster 3 (C3) and contributors greater than 10 in Cluster 4 (C4).

\textbf{SoCCMiner:} We employ SoCCMiner\cite{sridharan2022soccminer} to extract the contextual information of source code comment along with the comprehensive project attributes (including project, file, package, class, interface, enum, and, static block attributes). The essence of SoCCMiner is its ability to extract bi-directional contextual source code of a 
comment. For example: consider the comment ``TODO elastic?'' in Table \ref{tab:comparison_to_prior}. 
The unique capability of SoCCMiner apart from fetching parent trace for code comments is that it fetches both preceding and succeeding code context. It senses one of the special case in which the comment is in the first line of a function, and therefore to fetch the preceding code context, it fetches the entire function. Such bi-directional code context is crucial for code comprehension, comment generation of non-header comments.

\begin{table}[htbp]
\caption{Code Context Extraction Comparison}
\label{tab:comparison_to_prior}
\begin{tabular}{l|l}
\hline
\textbf{\begin{tabular}[c]{@{}l@{}}Context in prior work \cite{gelman2019source}\end{tabular}} &
  \textbf{\begin{tabular}[c]{@{}l@{}}Code Context in SoCCMiner\cite{sridharan2022soccminer}\end{tabular}} \\ \hline
                   &                                               \\
                   & \textbf{public void NoEvent(int ix, V val)\{} \\
// TODO elastic?   & // TODO elastic?                              \\
rows.add(ix, val); & rows.add(ix, val);                            \\
                   & \textbf{\}}                                  
\end{tabular}
\end{table}

%


\textbf{NLoN:} 
Source code comments often contain commented out source code, or other non-natural language elements. 
To remove code comments that do not contain natural language text, we utilize NLoN~\cite{mantyla2018natural} tool that accept code comments and perform a binary classification if the comment is natural language (`NL') or not (`Not'). The results are updated in PENTACET database for every comment under `NLON\_STATUS' column of the COMMENT\_ATTR table.

\textbf{SATD Feature Extrapolation:} 
We generate new SATD features based on the existing 64 established SATD features in \cite{potdar2014exploratory} with Sense2Vec~\cite{trask2015sense2vec}. 
Sense2Vec captures contextually similar words with its deep word embedding. For example, for feature `problematic', Sense2Vec returns multiple semantically similar words such as `troublesome', `undesirable', `contentious', `separate issue', etc., We use these extrapolated features (words) to evaluate our code comments for SATD. 

\begin{table}[htbp]
\caption{PENTACET (5C Dataset) Stats}
\label{tab:data_stats}
\footnotesize
\centering
\resizebox{\linewidth}{!}{%
\begin{tabular}{>{\centering\hspace{0pt}}m{0.156\linewidth}>{\centering\hspace{0pt}}m{0.152\linewidth}>{\centering\hspace{0pt}}m{0.165\linewidth}>{\centering\hspace{0pt}}m{0.165\linewidth}>{\centering\hspace{0pt}}m{0.165\linewidth}|>{\centering\arraybackslash\hspace{0pt}}m{0.133\linewidth}} 
\toprule
 & \textbf{CLUSTER }\par{}\textbf{ 1}\par{}\textbf{ (\#1 }\par{}\textbf{ Contributor)} & \textbf{CLUSTER }\par{}\textbf{ 2}\par{}\textbf{ (\#2-3 }\par{}\textbf{ Contributors)} & \textbf{CLUSTER }\par{}\textbf{ 3}\par{}\textbf{ (\#4-10 }\par{}\textbf{ Contributors)} & \textbf{CLUSTER }\par{}\textbf{ 4}\par{}\textbf{ (\textgreater{} \#10 }\par{}\textbf{ Contributors)} & \textbf{TOTAL} \\ 
\midrule
\textbf{\# Projects} & 1,576 & 2,202 & 3,004 & 2,314 & \textbf{9,096} \\
\hline
\textbf{\# Comments} & 1,055,517 & 2,498,870 & 4,660,481 & 15,079,741 & \textbf{23,294,609} \\ 
\hline
\textbf{\# Invalid Comments} & 31,118 & 37,778 & 84,876 & 152,193 & \textbf{305,965} \\ 
\hline
\textbf{\# NL Comments} & 733,409 & 1,737,444 & 3,281,562 & 10,336,159 & \textbf{16,088,574} \\ 
\hline
\textbf{\# SATD  Comments} & 23,039 & 54,052 & 73,204 & 329,506 & \textbf{479,801} \\ 
\hline
\textbf{\% SATD }\par{}\textbf{Comments} & 3.14 & 3.11 & 2.23 & 3.18 & \textbf{2.98} \\
\bottomrule
\end{tabular}
}
\end{table}

Sense2Vec extrapolated `Easy to Find' features such as (`sabotaging', `damaging', `counter-productive', etc.,) are passed to the first round of automated SATD annotation. The first author manually evaluates the results by reading the comments. The well-established `Easy to Find' \cite{yu2020identifying} features such as `TODO', `FIXME', and `XXX' were excluded in the manual validation. During the validation process, the first author creates the heuristics for `Hard to Find'~\cite{yu2020identifying} SATD. For example, the `Hard to Find' SATD feature `not sure if this is' returns results as the following:

\setlength{\fboxrule}{0.5mm}
\fbox
{
\begin{minipage}{0.4\textwidth}
    \textbf{//not sure if this is really what should be done}

    \textbf{//not sure if this is the right thing to do.}

    \textbf{//not sure if this is needed?}

    \textbf{//not sure if this is correct, double check!!}

    \textbf{//not sure if this is required, but looks feasible}
\end{minipage}
}
\vspace{3pt}


This iterative process involved multiple iterations of annotation, manual evaluation,  creating heuristics (pattern matching), and creating features for the `Hard to Find' SATD.

\textbf{SATD Annotation:}
Features that attract too many false positives are removed from the SATD feature list. For example, `cause problem' will capture the comment `// Empty bboxes can cause problems' which sound right at first. Later, after manual scanning, several header comments had the following comment content `//This won't cause problems'. Such features are removed from the SATD feature list to avoid overfitting features. 
We choose C1 cluster for heuristics' creation since we wanted to start with the smallest (in terms of \#comments) considering the time and manual effort involved in the annotation. It took two months for the first author to complete the SATD feature annotation. We have included both `Easy to Find' and `Hard to Find' SATD features in \cite{pentacet23}. Table \ref{tab:data_stats} contains the SATD split for each cluster.
To the best of our knowledge, this is the first attempt to explore and annotate the `Hard to Find' SATD. We have characterized 655 `Hard to Find' SATD features in PENTACET\cite{pentacet23}. Although this process is semi-automated, involving a human annotator, we believe this data will be the foundation for much more sophisticated, completely automated techniques for `Hard to Find' SATD detection.

For evaluating the quality of the dataset, we obtained a stratified random (balanced sample strength to accommodate both `SATD' and `Non-SATD') and a statistically significant sample size 385\footnote[5]{https://www.surveymonkey.com/mp/sample-size-calculator} accounting for 95\% confidence level with 5\% margin for error. The first author manually validated 1,540 samples. To remove any bias in the annotation process, we took a random sample of 97 from each cluster, and the second author annotated a total sample size of 388. The second author is a doctoral researcher whose primary area of research is Technical Debt.
The Inter-Rater Agreement (IRA) is evaluated through Cohen's Kappa Co-efficient. The IRA for our dataset achieved k=0.75, which is substantial \cite{fleiss1981measurement,emam1999benchmarking}. 

\section{Data Model}
\label{dm}

\begin{figure}[h]
  \centering
  \includegraphics[width=0.45\textwidth]{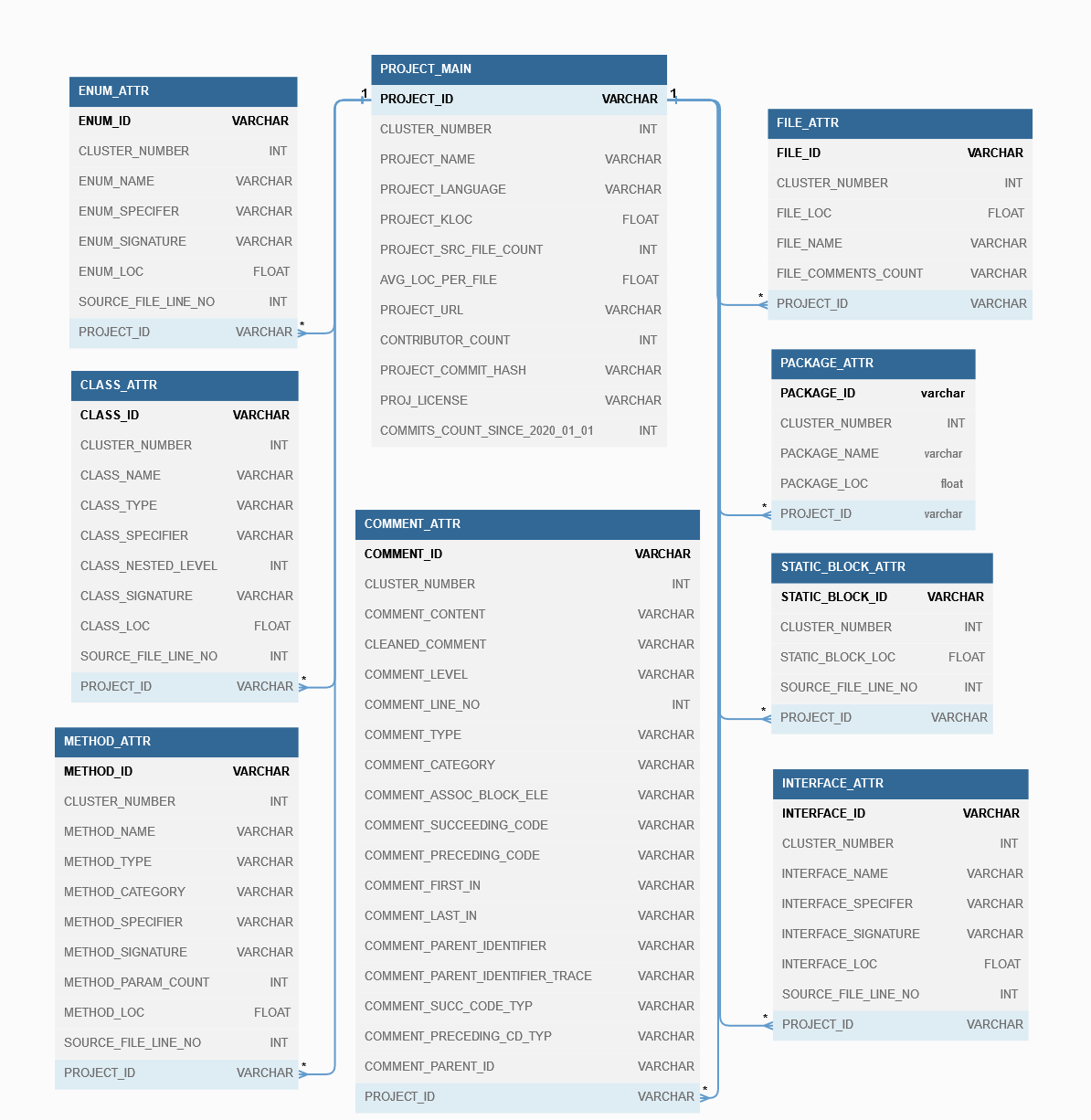}
  \caption{PENTACET Schema}
  \label{schema}
\end{figure}

We use PostgreSQL\footnote[6]{https://www.postgresql.org/} 14.0 database to store the PENTACET schema. All the project meta information are stored in the `PROJECT\_MAIN' table as observed in Figure \ref{schema}. 

The database dump file for the PENTACET contains more than 20 million bi-directional contextual code comments along with comprehensive project attributes as in \cite{sridharan2022soccminer}. 


The attributes of each granularity in a project are stored in their respective attribute table. For example, file attributes are stored in `FILE\_ATTR' table, method attributes are stored in `METHOD\_ATTR' table, etc., The main table containing a bulk of data is `COMMENT\_ATTR' table. The `COMMENT\_ATTR' table contains `NLON\_STATUS' column that contains either 'NL' (Natural Language) or `Not' (Not Natural Language) or `INVALID' (empty comments after removing any decorative comment symbols such as `///',`---',`***',`\#\#\#', and trimming trailing and leading whitespaces). The `SATD\_AFFLICTION' column contains the SATD status of the comment and `SATD\_FEATURE' column contains the SATD features that caused the comment to be SATD in the annotation process.

For illustration purposes,
the query i) fetches the SATD afflicted natural language comments from those projects whose Lines of Code (LOC) is greater than 10,000 and can be from clusters 1 to 4 based on the input.

\setlength{\fboxrule}{0.5mm}
\fboxsep=1pt



  





\begin{center}
\footnotesize
\begin{minipage}{0.42\textwidth}
  \textbf{i) }SELECT COMMENT\_ID, COMMENT\_CONTENT 

FROM COMMENT\_ATTR 

  WHERE CLUSTER\_NUMBER=[1$-$4] AND
  
  PROJECT\_ID IN 
		(SELECT PROJECT\_ID 
  
  FROM PROJECT\_MAIN WHERE PROJECT\_LOC $>$ 10000)

		AND NLON\_STATUS = `NL'

		AND SATD\_AFFLICTION= `SATD'
\end{minipage}
\end{center}
\vspace{3pt} 

Similarly, comments can be fetched based on the other granular (class, method, interface, etc.,) attributes.

\section{Applications}
\textbf{Modeling:}
Modeling code comments with more than 16 M natural language comments with code contextual information and almost 0.5 M SATD comments will aid deep learning for source code comments and technical debt research. The extended attributes of all project granularities associated with the contextual code information enable large-scale source code characterization for SATD.

\textbf{Predictive Analytics:}
The characterization of SATD in conjunction with project size (in LOC), file size (in LOC), other granular code constructs' (such as class, method, interface, etc.,) size (in LOC), and other attributes generated by SoCCMiner \cite{sridharan2022soccminer}
can help investigate its relation with SATD enabling correlational analysis and forecasting applications.

\textbf{Behavioural Analytics:}
The contributor mapping to the contextual source code metrics can help investigate developers' commenting practices across different project sizes, e.g., are there more or fewer comments in large projects? It can further help study sentiment analysis and assist with correlation analysis of developers' commenting practices with SATD. 
\section{Ethical Considerations}
We carefully considered ensuring we adhere to the ethical mining guidelines as in \cite{gold2022ethics}. In particular, repositories with no license attached to them were ignored and not included in the dataset. We carefully monitored to anonymize any personal information (author\_id of the repository) while processing PENTACET data.
\section{Threats to Validity}
\textbf{Internal Validity:} The primary limitation of this dataset is that the SATD patterns identified during the iterative process are taken from cluster 1 (C1) alone. 
However, this is a step towards capturing the fine-grained features of SATD that can enable deep learning techniques to extract such data from unseen data. 
The comments might contain noise, even in comments categorized as natural language. Data is prone to be noisy; careful curation and initial efforts toward automating noise removal are paramount.

\textbf{External Validity:}  
Although this is huge, open source data, it is difficult for this dataset to characterize SATD features exhaustively or claim generalization of SATD features across all the projects. However, this work is the largest SATD dataset to our knowledge.

\section{Related Work}
The Code and Comments dataset \cite{gelman2019source} and CoDesc\cite{hasan2021codesc} dataset captures code comment pairs. 
Comparatively our data is larger and in addition, SoCCminer supplements bidirectional code context information at all granular level (class, method, interface, etc.,).   
Such context information will be very crucial for identifying the locale or code context of non-header comments. The Technical Debt dataset \cite{lenarduzzi2019technical} primarily concentrates on TD generated by SonarQube\cite{sonarqube23},
while PENTACET characterizes `Hard to Find' SATD present in source code comments. 20-MAD \cite{claes202020} dataset focuses on comments from issues and commits while our dataset focuses on source code comments along with source code contextual information.

\section{Conclusion}
We apply extensive curation for mining OSS repositories and present the largest bi-directional code contextual data of 23 M source code comments at all source code granularities for 9096 Java projects. Further, we fetch and associate each repository's contributor count, enabling correlational and characterization analysis relating to source code comments and developer team size. Thirdly, we have compiled the largest SATD dataset through a semi-automated annotation process after removing non-natural language content (source code) and noise (empty comments or dummy comments with just symbols) using the NLoN tool. In the process, we extended the SATD features from \cite{potdar2014exploratory} with Sense2Vec\cite{trask2015sense2vec} and added manually verified features that detect `Hard to Find' SATD as suggested in \cite{yu2020identifying}. We achieved a substantial agreement in the inter-rater agreement during manual validation. 

For future work, 
we plan to fetch the source code, static analysis, git process metrics for all the repositories in PENTACET and associate SATD comments with the generated metrics using the unique combination of comment parent identifier and comment trace (SoCCMiner\cite{sridharan2022soccminer} attributes). It will enable software researchers to investigate the correlational study and improve SATD forecasting.

\bibliography{conference_101719}
\end{document}